# La relatividad en la cultura de la Argentina de inicios del siglo veinte

*Alejandro Gangui* [*], *Eduardo L. Ortiz*[†]

**Introducción**

En trabajos anteriores hemos considerado diversos aspectos de la recepción de las teorías de Albert Einstein en nuestro país. En particular, discutimos el impacto de su visita a la Argentina en 1925, y el contenido preciso de las conferencias y publicaciones científicas que Einstein comunicó mientras estuvo aquí. Hemos discutido también, críticamente, diferentes documentos sobre temas afines a la teoría de la relatividad publicados en el país durante las primeras décadas del siglo pasado, en los que el nivel técnico fue sumamente variado. Finalmente, nos hemos ocupado de un documento olvidado, que Einstein preparó para su visita pero que, una vez en el país, creyó conveniente no comunicar (Ortiz, 1995, Gangui y Ortiz, 2005, 2008, 2009, 2011, 2014).

En el presente trabajo, que es parte de esa misma línea de investigaciones, nos proponemos considerar la carrera científica, cultural y pedagógica de Enrique Loedel Palumbo (1901-1962), que fue uno de los principales actores locales durante la visita de Einstein a la Argentina. Asimismo, fue uno de los fundadores de la investigación en temas de relatividad en el Río de la Plata. Además de sus intereses estrictamente científicos, Loedel Palumbo se ocupó seriamente de la pedagogía de las ciencias y se interesó por problemas filosóficos, culturales y artísticos de su tiempo.

**Sus estudios en La Plata: Richard Gans**

Enrique Loedel Palumbo (Loedel, a partir de ahora) nació en Montevideo el 29 de junio de 1901 y cursó la totalidad de sus estudios primarios y secundarios en el Uruguay. Luego de terminar su bachillerato de cuatro años, cursó los dos años de estudios "Preparatorios" de Ingeniería en la Universidad de la República de Montevideo. Al finalizarlos, se trasladó a la Universidad de La Plata (UNLP) donde se inscribió en la Licenciatura en Ciencias Físico-Matemáticas en la Facultad de Ciencias Físico-Matemáticas (Loedel Palumbo, o LP, 1940b, p. 2). En ese momento esta universidad tenía el laboratorio de física más avanzado y el personal mejor entrenado en esa disciplina de toda América Latina.

Loedel tenía también un interés especial por la educación, que lo llevó a cursar, conjuntamente, el Profesorado en Matemáticas y Física en la Facultad de Humanidades de la UNLP. En diciembre de 1923 recibió su diploma en esta última Facultad, lo que le permitía el acceso a la enseñanza secundaria. Esta posibilidad tuvo, más adelante, una importancia considerable en su vida profesional.

En 1925 completó la Licenciatura en Física y, paralelamente a sus estudios, preparó su tesis de doctorado. El entonces director del Instituto de Física, Richard Gans, que era una figura de nivel internacional y sin duda alguna el científico más destacado de ese instituto, fue su director de tesis.

---

[*] Universidad de Buenos Aires, CONICET

[†] Imperial College, Londres

En su trabajo Loedel utilizó recursos ópticos -la espectroscopía y otras técnicas de laboratorio- como herramientas para el estudio de las propiedades ópticas y eléctricas de la sacarosa. El 14 de diciembre de 1925 defendió su tesis, y poco más tarde le fue otorgado el diploma de Doctor en Física.

Un resumen de la tesis, presentado para su publicación a principios de 1926, apareció impreso en la revista *Contribución al Estudio de las Ciencias Físicas y Matemáticas* (LP, 1926a). Esta fue la primera revista argentina de investigaciones originales en las ciencias físicas, fundada por Gans en la UNLP en 1913.

En esos años Gans estaba haciendo un esfuerzo serio por educar a sus alumnos, no solamente a hacer investigación original, sino también a concretar sus resultados y presentarlos en una forma publicable tanto en el país como en la supuestamente más exigente prensa científica internacional. Una versión más reducida de la tesis de doctorado de Loedel fue publicada en la prestigiosa revista alemana *Annalen der Physik* con un apéndice teórico de Gans (LP, 1926b). Lamentablemente, Gans dejó la Universidad de La Plata para hacerse cargo de una cátedra en Königsberg en 1925: Loedel fue su último discípulo en La Plata.

**Loedel Palumbo y la visita de Einstein a la Argentina**
En 1925 Albert Einstein visitó la Argentina y dictó una serie de conferencias sobre su teoría de la relatividad en la Universidad de Buenos Aires (UBA); además dictó conferencias especiales en las universidades de La Plata y Córdoba. En esas semanas la Academia Nacional de Ciencias Exactas, Físicas y Naturales de Buenos Aires organizó una reunión especial en su homenaje, en la que le ofrecería el diploma de Académico Honorario.

Esa sesión especial tuvo lugar el 16 de abril y fue coordinada por su entonces presidente, el naturalista Dr. Eduardo L. Holmberg, uno de los más destacados científicos de la Argentina de esa época, con la colaboración del Ing. Nicolás Besio Moreno.

Varios académicos y especialistas en temas de física, fueron invitados a proponer al homenajeado preguntas sobre su teoría. Ellos fueron los académicos Dr. Ramón G. Loyarte y Dr. Horacio Damianovich y tres jóvenes investigadores: los físicos Dr. Teófilo Isnardi, Dr. José B. Collo y el astrónomo Ing. Félix Aguilar. Los tres últimos acababan de publicar una detallada exposición sobre la teoría de la relatividad, que habían desarrollado previamente en un ciclo organizado por la Sociedad Científica Argentina en 1922-23, precisamente en preparación de la visita de Einstein. Este trabajo ha sido analizado en detalle en (Gangui y Ortiz, 2011).

También fue invitado a participar en la sesión de preguntas el joven estudiante de La Plata Enrique Loedel Palumbo, que no estaba aún doctorado, ni siquiera graduado. Tanto las preguntas, como las supuestas respuestas de Einstein, fueron publicadas en los *Anales* de la Academia (Academia, 1928). Aunque las preguntas de Isnardi, Collo y Aguilar revelan un conocimiento detallado de la teoría, la intervención de Loedel, muy específica, sugiere un interlocutor que se encontraba investigando activamente en el campo de la teoría de la relatividad.

La pregunta que formuló el joven Loedel hizo que el visitante se detuviera a responderla con cierta atención. Si las minutas de esa reunión pueden tomarse como una referencia confiable, Einstein habría encontrado interesante el problema sobre el que Loedel le preguntaba. Este era: "¿es posible hallar una representación de la superficie espacio-tiempo de dos dimensiones en un espacio euclídeo de tres?" (Academia, 1928, p. 330). Este problema supuestamente había conducido al joven a formular un sistema de ecuaciones diferenciales parciales no-lineales que no había logrado resolver. Según esa misma publicación, "el profesor Einstein contesta que [esas ecuaciones] no han sido resueltas, y que el problema de investigar la forma de la superficie espacio-tiempo sería muy interesante" (Ibid., p. 331).

Un año más tarde, hacia mayo de 1926, Loedel consiguió resolver el problema que había planteado a Einstein y publicó su solución, primero en *Contribución al Estudio de las Ciencias Físicas y Matemáticas* (LP, 1926c) y, en el mismo año, en *Physikalische Zeitschrift* (LP, 1926d). La suya es una nota breve, no ya un trabajo fundamental, pero el problema resuelto no era necesariamente simple.

Puede decirse que esa nota breve de Loedel de 1926 abrió dos capítulos nuevos para la física en la Argentina: el de la investigación original en física teórica y el de la teoría de la relatividad como tema de investigación, no ya de divulgación popular o avanzada como había sido hasta entonces.

En junio, Loedel volvió a ocuparse de temas de relatividad en un trabajo sobre la velocidad de la luz en un campo gravitacional (LP, 1926e). Sin embargo, luego de la partida de Einstein el interés por esa teoría quedó reducido a las tareas de muy pocos cultores de la física y principalmente con propósitos didácticos; gradualmente se fue extinguiendo: tanto en la Argentina como fuera de ella, el interés comenzó a centrase en la nueva mecánica cuántica.

En su actividad docente durante esos años Loedel alternó su trabajo en la Facultad de Ciencias de UNLP y en la Facultad de Humanidades de esa misma universidad con tareas de enseñanza en escuelas secundarias. Luego de su primera designación como profesor en el Colegio Nacional de La Plata, en abril 1924, fue nombrado Jefe de Trabajos Prácticos de Física en la Facultad de Ciencias en 1926 y en 1927 adquirió una segunda cátedra en el mismo Colegio. También en 1927 fue designado profesor suplente de Física General (A ó B) en la Facultad de Ciencias y profesor suplente de Geografía Matemática en la de Humanidades. Más tarde, en 1936 agregó a esas tareas la de profesor en el Liceo de Señoritas, también en La Plata (LP, 1940b, p. 2).

Su talento había sido detectado tempranamente también en la Universidad de Buenos Aires, donde se lo incorporó como Conservador del Gabinete de Física aun antes de la visita de Einstein. Quizás esta no fuera una designación ideal para un físico teórico pero, muy posiblemente, era la única disponible dentro del presupuesto de esa institución. Lamentablemente, esta conexión no sobrevivió su viaje a Alemania, al que nos referiremos a continuación.

**La década de 1930, física y filosofía de la ciencia**
En 1928 y 1929 Loedel obtuvo una beca de la UNLP para trasladarse a la Universidad de Berlín, donde hizo estudios de física bajo la dirección de dos distinguidos maestros: Max Planck y Erwin Schrödinger (LP, 1940b, p. 2) indicando también, más tarde, que en ese período estuvo en relación con Einstein, lo que no es sorprendente, y con Hans Reichenbach (LP, 1959, p. 3).

A través de este último tomó un contacto directo con el movimiento positivista lógico que buscaba una fundamentación más precisa de las ciencias exactas. En 1940 Loedel (LP, 1940b, p. 1) indicó que "en algunas épocas de mi vida he estudiado filosofía con fervor, en mi juventud fui racionalista cartesiano, más tarde escéptico y por el momento el empirismo consecuente del círculo de Viena cuenta con mis simpatías".

En las décadas de 1930 y 1940 las contribuciones de Loedel reflejan un interés profundo por la discusión de los fundamentos de las ciencias físicas, a la que contribuyó con varios aportes. En 1935 regresó a Europa para participar en una reunión del primer *Congrès International de Philosophie Scientifique,* reunido en la Sorbonne, Paris, en el que participaron, entre otros, A. J. Ayer, R. Carnap, F. Enriques, Ph. Frank, A. Lautmann, A. Tarski, y H. Reichenbach, éste último ya emigrado a Turquía.

La enseñanza de la física a un nivel elemental, es decir, la comunicación precisa de los resultados de esa ciencia a la juventud, y el análisis de ese proceso, fue una preocupación consecuente con sus estudios en la Facultad de Humanidades a principios de la década de 1920. A este último interés puede, quizás, atribuirse el esfuerzo considerable que dedicó a la autoría de una serie de obras de enseñanza de la física y de la matemática a nivel de la enseñanza secundaria. Esas obras dejaron una marca por su originalidad y por la búsqueda de una fundamentación racional.

Después de su regreso de Alemania, Loedel continuó en contacto con las dos Facultades de la UNLP y, al mismo tiempo interrumpió su colaboración docente con la UBA, aunque, como veremos en seguida, tuvo contacto con otras instituciones de alto nivel de esa ciudad. Es posible que su contacto con la UBA fuera rescindido por razones de economía del tiempo; sin embargo su colaboración con esa universidad hubiera sido oportuna en esos años.

Poco después de su regreso de Alemania, a principios de la década de 1930, Loedel fue invitado a dictar un ciclo de conferencias en el recién establecido Colegio Libre de Estudios Superiores (CLES) de Buenos Aires. En las dos décadas siguientes el CLES se convertiría en una de las más altas tribunas de estudios avanzados de la Argentina, una alternativa a la Universidad oficial a la que se comenzó a percibir como excesivamente concentrada en la formación de profesionales.

En 1930, en un primer ciclo en el CLES, Loedel dictó nueve conferencias sobre la estructura del átomo; el año siguiente fue invitado nuevamente por el CLES para dictar otras seis conferencias sobre el mismo tema. Más tarde, resúmenes de esas y de otras conferencias suyas fueron publicados en *Cursos y Conferencias*, la revista oficial del CLES (LP, 1931, 1934a).

**La enseñanza de las ciencias a la juventud**

En la víspera de su partida a Alemania, Loedel había colaborado con otro físico de La Plata, el ya mencionado Loyarte, en un exitoso *Tratado Elemental de Física*, cuyo primer volumen apareció en 1928 y un segundo en 1932 (Loyarte y Loedel Palumbo, 1928, 1932).[1] Esta obra alcanzó justa popularidad, al punto de seguir republicándose por más de un cuarto de siglo.

Separadamente, ambos físicos continuaron preocupándose por la educación en el área de sus intereses. En el caso de Loyarte sus obras de texto dieron preferencia al nivel universitario. Otros físicos, como Isnardi y Collo, miembros del mismo grupo inicial de la UNLP, escribieron también textos adaptados a los programas de la UBA o de otros centros: la Escuela Naval fue uno de ellos. El éxito de esas obras sugiriere que la modernización de la enseñanza de la física a nivel secundario y universitario, es decir, su aporte a la elevación del nivel cultural del país, resultó ser –finalmente– un resultado concreto e importante del esfuerzo modernizador iniciado por su organizador, Joaquín V. González, en la UNLP con la asistencia específica de Emil Bose, y luego de Gans, en el caso de la física.

Con posterioridad a aquella obra con Loyarte, y más particularmente en la primera mitad de la década de 1940, Loedel escribió unas 15 obras de texto para la enseñanza secundaria de la matemática y la física, algunas de ellas en colaboración con Salvador de Luca, a los que dio un enfoque muy personal.

Esas obras fueron publicadas por la editorial Ángel Estrada, uno de los editores más cotizados en el área de las obras de texto elementales en ese momento. En ellas Loedel trató de dar una perspectiva más amplia que la que requerían los programas de estudio, agregando notas históricas y perspectivas modernas que no siempre estuvieron presentes en otras obras de esa época.

Entre esas obras se destaca el texto de *Física elemental* (LP, 1940a) que fue recibido con satisfacción por la crítica local, señalándose en una de ellas (Textos, 1940) que el autor no había escatimado esfuerzo "en su propósito didáctico" y que además de su mérito didáctico había mostrado el "lugar que corresponde al conocimiento de la física en el acervo cultural". También se destaca su *Cosmografía o Elementos de Astronomía*, un texto de más de seiscientas páginas redactado, nuevamente en colaboración con de Luca (Loedel Palumbo y de Luca, 1940). La primera edición apareció en 1940 y en la década siguiente fue reeditado nuevamente en casi cada año. A partir de 1941 comenzó a aparecer una serie de textos sobre la Geometría, la Aritmética y el Álgebra elementales (Loedel Palumbo y de Luca, 1942), redactadas con el mismo colaborador.

En 1949 Loedel publicó un interesante estudio crítico sobre la enseñanza de la física (LP, 1949a); en sus 15 capítulos se ocupó sucesivamente del método de la enseñanza, la experimentación, la interconexión de los conceptos científicos y el significado de una teoría física; consideró luego al alumno de física y analizó más adelante diversos conceptos centrales de la física: masa, temperatura, causalidad; finalmente consideró el papel que puede jugar la historia de la física y los recursos didácticos abiertos al profesor. Ese libro fue publicado dentro de una nueva serie, sobre "Ciencias de la Educación", iniciada por la Editorial Kapelusz, de Buenos Aires; ésta era una nueva empresa editorial con recursos técnicos que permitían hacer un uso muy amplio de las ilustraciones.

En total Loedel escribió más de cuatro mil páginas de obras de texto en la década de 1940 en las que ejemplificó sus ideas sobre las ciencias exactas y sobre su enseñanza.

El interés de Loedel por los fundamentos y la didáctica de las ciencias fue compartido con otros científicos argentinos de ese período y se puede asociar con diversos avances institucionales. En particular con la creación de un capítulo local de la *Académie Internationale d'Histoire des Sciences* en Buenos Aires, con el apoyo de Julio Rey Pastor y Aldo Mieli (Ortiz, 2012, pp. 425-27); hemos señalado más atrás su participación en el primer congreso de filosofía de la ciencia, en Paris.

Loedel se preocupó también por las prácticas de laboratorio en la enseñanza de la física en los colegios de La Plata; en su expediente del Colegio de Señoritas de esa ciudad hay referencias a su donación de instrumentos y, también, a una nota oficial (LP, 1940b, p. 3), fechada en 1942, en la que se le agradecía su contribución al desarrollo del laboratorio de ese colegio.

En 1934, entre medio de su obra en colaboración con Loyarte (y de la mencionada controversia) y su segunda larga serie de textos, Loedel publicó un libro de poesías titulado *Versos de un físico; física y razón vitaI* (LP, 1934b), que fue favorablemente recibido por la crítica literaria de los principales periódicos locales. Sobre este libro de poesías escribió su autor que lo consideraba "muy bueno" (LP, 1940b, p. 5). Loedel ha dicho que, en poesía, se consideraba discípulo de Rubén Darío, habiendo sido un admirador de Becquer en su primera juventud (Ibid., p. 1).

**Loedel Palumbo:** *Física relativista*

Hacia mediados de la década de 1940 Loedel volvió a sus antiguas preocupaciones en el campo de la teoría de la relatividad. Su figura como investigador original en esa área, preocupado a la vez por la comprensión y difusión de las ideas científicas, conserva hoy un interés considerable dentro de los estudios sobre la relatividad en la Argentina. En particular, en esos años, Loedel hizo esfuerzos considerables para construir herramientas conceptuales capaces de permitir al estudioso avanzar más profundamente en el abordaje de esa teoría.

En 1943, luego de un golpe militar, las universidades argentinas sufrieron una "depuración" política que afectó a un grupo extenso de profesores. Loedel se contó también entre los cesantes.[2] Meses más tarde, la intensificación de la oposición al régimen militar forzó una reconsideración de su actitud y la reincorporación de los afectados. En 1946 se repitieron, aunque en una dimensión mucho mayor, las expulsiones masivas por razones políticas. En marzo de 1947 Loedel fue dejado cesante en sus cargos, tanto en la UNLP como en los establecimientos de enseñanza secundaria anexos a ella. Recién en 1956 sería reincorporado nuevamente a aquella universidad.

En junio de 1948, Loedel fue invitado por la Universidad Nacional de Cuyo (UNCu) a incorporarse como profesor contratado en la Facultad de Ingeniería y Ciencias Exactas, Físicas y Naturales que ésta sostenía en San Juan (LP, 1959, p.7).

La presencia de un buen número de especialistas de calidad en su plantel docente (muchos de ellos cesanteados de otras instituciones) hizo posible que la UNCu pudiera considerar la posibilidad de desarrollar un interesante y ambicioso proyecto de investigación científica. Se organizó un instituto que tomó el nombre de Departamento de Investigaciones Científicas, o DIC. Como el cuerpo central de la universidad, el DIC fue establecido en la ciudad de Mendoza, con ramas fuera de ella. Loedel fue miembro de su primer Consejo Técnico colaborando también con otro organismo de esa misma universidad, el Centro de Estudios Físico Matemáticos, creado el 27 de junio de 1948.

También en 1948 Loedel participó en un Congreso de Filosofía que se reunió en Mendoza entre el 30 de marzo y el 9 de abril y que fue el primero de carácter internacional en esa disciplina en la Argentina. Allí leyó una nota sobre el problema de las magnitudes físicas (LP, 1949b). En esta contribución consideró críticamente el concepto de magnitud, estudiando el impacto de la teoría de la relatividad sobre dicho concepto. Por otra parte, en una reunión del Centro de Estudios Físico Matemáticos de la UNCu celebrada en Mendoza en 1948, Loedel se ocupó del problema de la aberración de la luz en la teoría de la relatividad, enviando un trabajo sobre el mismo tema para su publicación en los Anales de la Sociedad Científica Argentina (LP, 1948). Este interesante trabajo daría lugar a una propuesta alternativa a los diagramas de Minkowski, que discutiremos en detalle en otra ocasión.

Su contrato caducó a fines de 1953 y, en sus propias palabras, "[n]uevamente quedó totalmente libre de toda tarea docente en 1954 y 1955" (LP, 1959, p. 7). En esos años publicó varios trabajos y redactó su libro *Física relativista*, publicado en diciembre de 1955 (según LP, 1957, p. 78).

El libro *Física relativista* (LP, 1955) apareció nuevamente bajo el sello de la Editorial Kapelusz, de Buenos Aires; en esa obra volvió a ocuparse de la teoría de la relatividad y, según explica el autor, era su contribución personal a la celebración del cincuentenario de la aparición del primer trabajo de Einstein sobre esa teoría. Esta es una obra extensa, accesible a un público con conocimientos universitarios básicos, en la que el autor ofrece un estudio histórico crítico de la teoría de la relatividad y presenta algunas novedades.

Su obra fue recibida favorablemente por los críticos locales; uno de ellos, el matemático e historiador de la ciencia Ing. José Babini expresó que se trataba de "un excelente y útil tratado para el conocimiento y comprensión" de una parte importante de las conquistas de la física del siglo 20 (Babini, 1956). Sin embargo, otros consideraron que se trataba de una presentación muy personal de la teoría: es posible que sea precisamente en este punto donde resida el principal interés de esa obra.

Como indicamos más atrás, en noviembre de 1956, casi una década después de haber quedado cesante, fue reincorporado a su cátedra en la UNLP como profesor titular de Física I y II y de Física Teórica I; poco más tarde fue designado vicepresidente del Consejo de Investigaciones Científicas de la Provincia de Buenos Aires. Asimismo fue Jefe del Departamento de Física y Matemáticas del Colegio Nacional de La Plata y miembro correspondiente de la Academia Nacional de Ciencias de Lima. Loedel Palumbo falleció en la ciudad de La Plata el 31 de julio de 1962.

**Notas**

1. Con este mismo colega, quien había llegado a ser el Rector de la UNLP en 1928, Loedel tuvo una seria controversia científica, lo que seguramente no le agilizó su promoción en los ámbitos académicos platenses del momento.

2. Fue dejado cesante el 1 de septiembre de 1943 (LP, 1940b, p. 8).